# Identifying the pairing mechanism in Fe-As based superconductors: gaps and isotope effects

A. Bussmann-Holder[1], A. Simon[1], H. Keller[2] and A. R. Bishop[3]

[1]Max-Planck-Institut für Festkörperforschung, Heisenbergstr. 1, D-70569 Stuttgart, Germany

[2]Physik Institut der Universität Zürich, Winterthurerstr. 190, CH-8057 Zürich, Switzerland

[3]Los Alamos National Laboratory, Los Alamos, NM87545, USA

The temperature dependencies of the coupled superconducting gaps, observed in Fe-As based superconducting compounds is calculated and a *universal* temperature scaling observed which is only present if the coupled order parameters both have s-wave symmetry. Predictions for possible isotope effects on the transition temperature are made if phonons are involved in the pairing or polaronic effects are of importance. Comparison to experimental data is given where these are available.

The unusually high superconducting transition temperatures of the newly discovered Fe-As based superconductors [1, 2], and their proximity to a magnetic ground state have been taken as evidence that the origin of superconductivity might be unconventional and related to magnetic fluctuations. Also, the observation of only small lattice responses at the onset of superconductivity [3, 4] together with numerical results [5] have prompted advocacy for some exotic non-phononic pairing mechanism [6, 7]. However, opposite to cuprates, a sharp line separates the magnetic phase from superconductivity which suggests that these phenomena are not related to each other. Another important finding is the observation of multiple gaps either detected directly through tunneling experiments [8] or indirectly through the temperature dependence of the superfluid density [9 – 14], where nodeless gaps with s-wave symmetry have been observed. Also, it was possible recently to determine the iron isotope effect on $T_c$ which falls within the BCS limit [15]. This latter finding – even though not being a proof of lattice involvement – is an indication that the lattice plays a non-negligible role for superconductivity.

Here we present theoretical results for a two-band superconductor with very specific predictions as to how the gaps evolve with temperature, demonstrate that a *universal* scaling of the gaps is present if both are of s-wave symmetry, and show how an isotope effect arises if conventional lattice vibrations are involved contrasted to polaronic effects which play a very different role for the isotope effects. The latter local lattice coupling has been shown to be relevant for unconventional isotope effects in cuprate superconductors [16] and is also realized in Fe-As based compounds, even though much weaker, as suggested from EXAFS experiments [17].

Multi-band superconductivity has been considered early on as an enhancement mechanism of $T_c$ [18]. Its discovery much later in Nb doped $SrTiO_3$ [19] seemed to be an exception for a long time. Before the discovery of $MgB_2$ [20] it was, however, pointed out that two-gap superconductivity with mixed order parameters is also realized in cuprates [21] since many experiments are not compatible with a single d-wave order parameter. This idea has been shown recently by various experimental techniques [22], to indeed be true.

The Fermi surface of Fe-As based superconductors is complex with diverse electron and hole type sheets which might give rise to multiple superconducting gaps [23, 24]. However, two flat bands have been identified along the Γ-Z and A-M high symmetry directions [25], which admit to reduce the complexity to a two gap problem, both with s-wave symmetry, since this seems to be indicated experimentally. The coupled gap relations are derived from an effective two-band extended BCS Hamiltonian with interband interactions. The bands are Fe d-As p hybridized. This hybridization is crucial since the magnetic properties of the parent compounds are strongly dependent on it [24]. In addition, it is assumed that a band with large density of states leading to a dominant gap is present which induces superconductivity via interband interactions in the second band with a small density of states. Such a scenario corresponds to almost localized states coupled via a highly dispersive band to another small density flat band. Within this model the coupled gap equations are given by [26]:

$$<c^+_{k_1\uparrow}c^+_{-k_1\downarrow}> = \frac{\overline{\Delta}_{k_1}}{2E_{k_1}}\tanh\frac{\beta E_{k_1}}{2} = \overline{\Delta}_{k_1}\Phi_{k_1}$$

$$<d^+_{k_2\uparrow}d^+_{-k_2\downarrow}> = \frac{\overline{\Delta}_{k_2}}{2E_{k_2}}\tanh\frac{\beta E_{k_2}}{2} = \overline{\Delta}_{k_2}\Phi_{k_2}$$

$$\overline{\Delta}_{k_1} = \sum_{k'_2}V_1(k_1,k'_1)\overline{\Delta}_{k'_1}\Phi_{k'_1} + \sum_{k_1}V_{1,2}(k_1,k_2)\overline{\Delta}_{k_2}\Phi_{k_2}$$

$$\overline{\Delta}_{k_2} = \sum_{k'_2}V_2(k_2,k'_2)\overline{\Delta}_{k'_2}\Phi_{k'_2} + \sum_{k_1}V_{2,1}(k_2,k_1)\overline{\Delta}_{k_1}\Phi_{k_1} \qquad (1)$$

with effective attractive intraband interaction $V_i$, $i=1,2$, and interband interaction $V_{1,2}=V_{2,1}$, $E_k = \sqrt{\tilde{\xi}^e_k + \overline{\Delta}^2_i}$, $\xi_k$ being the momentum $k$ dependent band energy and $\overline{\Delta}_i$ the superconducting gap in band $i$. We make the assumption $V_{1,2}=V_{2,1}$ in order to minimize the parameters. The coupled Equs. 1 are solved simultaneously and self-consistently for each temperature T, and $T_c$ is defined when $\overline{\Delta}_i = 0$. The nonlinear problem given by Eqs. 1 leads to both gaps going to zero simultaneously. Since, as outlined above, we start from the fact that one band has a substantially enhanced density of states as compared to the other one, we mainly determine $T_c$ through the intraband interaction in this band and hold it constant for all values of $T_c$ in the second band. Similarly, the interband interaction is taken to be independent of $T_c$. In this way we keep the problem as transparent as possible.

In Fig. 1a the coupled gaps are shown as a function of temperature for various values of $T_c$.

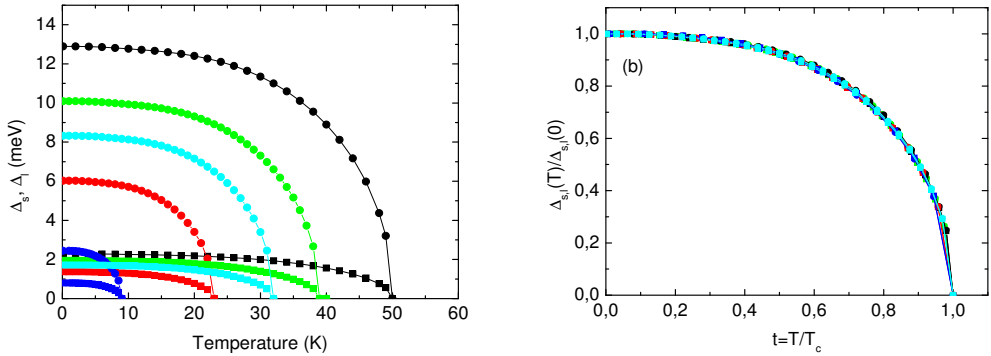

**Figure 1a** Temperature dependence of the small ($\Delta_s$) and the large ($\Delta_l$) gap for various values of $T_c$. **b** The normalized gap values displayed in **a** as functions of $t = T/T_c$.

In Fig. 1b the normalized gaps $\overline{\Delta}_i(T)/\overline{\Delta}_i(0)$ versus normalized temperature $t = T/T_c$ are shown. All values for the gaps, the small and large ones, fall on a single line. This scaling relation is not observed in systems where the coupled gaps have different order parameter symmetries [22]. It is thus possible to draw conclusions about the symmetry of the order parameters from the temperature dependence of the gap values. Note that, in the case of coupled d-wave order parameters, a linear T term should appear at low temperatures which is absent for s-wave gaps [28].

Besides the universal T-dependence of the normalized energy gaps, the gap to $T_c$ ratios exhibit interesting properties. For instance, for Al doped $MgB_2$ this ratio is substantially reduced for the smaller gap and enhanced for the larger one [27]. The average of both gaps is, however, in the BCS limit. This relation is again not fulfilled in a multiband superconductor with different pairing symmetries: Typically a d-wave order parameter leads to an appreciable enhancement over the BCS relation [28]. In Fig. 2a the gap to $T_c$ ratios and their average are shown for the scenario introduced above. A significant resemblance to Al doped $MgB_2$ exists, where the average ratio is again within the BCS value.

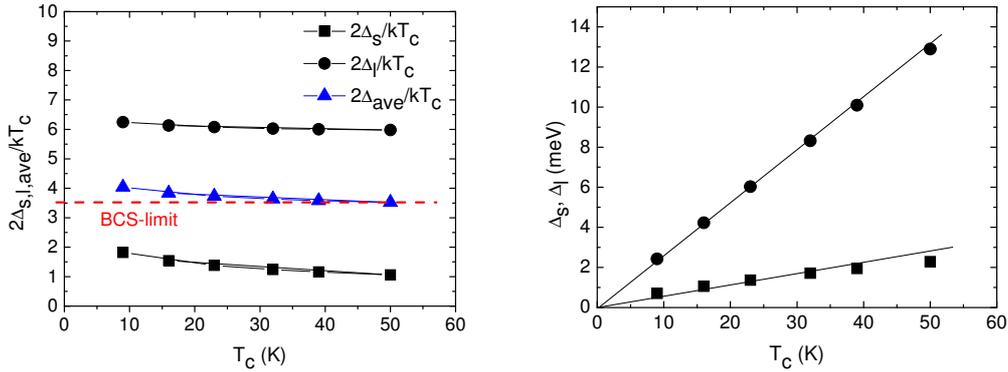

**Figure 2a** The individual and average gap to $T_c$ ratios as a function of $T_c$. **b** Energy gap values for the corresponding values of $T_c$.

However, a small but systematic dependence on $T_c$ is present: the ratio is slightly enhanced for small values of $T_c$, decreasing continuously to the BCS value with increasing $T_c$. In Fig. 2b the individual gap values are displayed as a function of $T_c$. It is important to emphasize again that only the intraband interaction in the band where the dominant gap opens, is varied with $T_c$ (see Table 1), whereas we hold the interaction in the band related to the smaller gap constant. In addition, also the interband interaction is held fixed for all investigated systems. In spite of these simplifications, obviously the smaller gap adopts a significant dependence on $T_c$ which implies that the coupling between gaps influences the smaller one to increase systematically with the larger one. The gap coupling together with the intraband coupling of the larger gap thus induces a substantial increase of the smaller gap with increasing $T_c$. The dependence of both gaps on $T_c$ deviates from linearity which is most pronounced for small values of $T_c$.

Next we explore possible origins of the Fe isotope effect on $T_c$ which has been reported to be almost in the BCS limit [15]. A variety of possibilities could explain any isotope effect within a two-band superconductor. E.g., for cuprates, the observed isotope effects have been shown to be a consequence of polaron formation; for a BCS superconductor they stem from electron-phonon interactions. In multiband superconductors lattice effects could be dominant in one channel only or become effective in both. On the other hand, a lattice mediated interband interaction alone could also cause an isotope effect. In the following we consider the possibilities that i) lattice mediated BCS superconductivity in the high density band is present; ii) polaronic couplings are present; and iii) the interband interaction is influenced by polaron formation. The two latter considerations are motivated by EXAFS studies [17], where an unusual upturn in the mean square Fe-As displacement evidences polaronic effects. We exclude that any dominant effects stem from the small density band, since the small density of states together with the small interband coupling will lead to a negligible isotope effect.

In Fig.3 we show the results for the three cases discussed above.

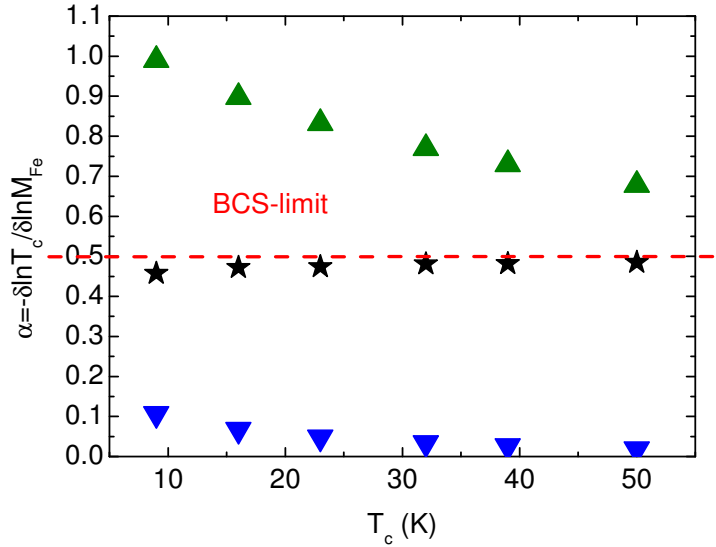

**Figure 3** The Fe isotope exponent $\alpha = -\delta \ln T_c / \delta \ln M_{FE}$ as a function of $T_c$. The upward triangles are the results for a polaronic coupling in the large gap channel, the downward triangles are the same but with polaronic coupling in the interband coupling channel. The stars are results for lattice mediated superconductivity in the dominant gap channel.

In the first case, namely lattice mediated BCS type pairing interactions in the dominant channel, the isotope effect, $\alpha = -\delta \ln T_c / \delta \ln M_{Fe}$, has a minor dependence on $T_c$, being around 0.45 for small values of $T_c$ to increase slightly to 0.48 for the maximum $T_c$ investigated here. In all cases it remains smaller than 0.5 but is always close to this value. This is actually what has been observed experimentally for $SmFeAsO_{1-x}F_x$ and $Ba_{1-x}K_xFe_2As_2$ [15]. Polaronic couplings in the large gap channel as well as in the interband interaction induce a strongly $T_c$ dependent α which is largest for small $T_c$ or small doping, respectively, to decrease with increasing doping or $T_c$. Such a doping dependence is present in cuprates [16, 22] and has been shown to stem from polaron formation. For Fe-As based systems this does not seem to be realized experimentally in spite of the fact that a small polaronic coupling exists. Since the first case, i.e., lattice mediated superconductivity in the large gap channel, is almost in quantitative agreement with experiment, we conclude that a rather conventional pairing mechanism is present.

**Table 1**

**Table captions**
The first column of the table shows the values of $T_c$, the second, third and fourth columns display the density of states times the interaction constants in the small, large and interband terms. The absolute gap values for the small and large gaps are given in the next two columns. The last three columns refer to $2\Delta/kT_c$ ratios for the small, the large and the average gap

| $T_c$ (K) | $N_l V_l(0)$ | $N_s V_s(0)$ | $N_{sl} V_{sl}(0)$ | $\Delta_l(0)$ (meV) | $\Delta_s(0)$ (meV) | $2\Delta_s/kT_c$ | $2\Delta_l/kT_c$ | $2\overline{\Delta}/kT_c$ |
|---|---|---|---|---|---|---|---|---|
| 9 | 0.208 | 0.018 | 0.06 | 2.423 | 0.707 | 1.823 | 6.247 | 4.035 |
| 16 | 0.242 | 0.018 | 0.06 | 4.229 | 1.063 | 1.541 | 6.133 | 3.837 |
| 23 | 0.268 | 0.018 | 0.06 | 6.025 | 1.370 | 1.38 | 6.08 | 3.73 |
| 32 | 0.297 | 0.018 | 0.06 | 8.316 | 1.713 | 1.242 | 6.030 | 3.636 |
| 39 | 0.317 | 0.018 | 0.06 | 10.089 | 1.949 | 1.16 | 6.002 | 3.581 |
| 50 | 0.346 | 0.018 | 0.06 | 12.895 | 2.285 | 1.06 | 5.983 | 3.522 |

We shortly comment on our choice of varying the intraband interaction in the high density of states band only. Experimentally, it is observed that $T_c$ strongly varies with chemical pressure, i.e., it first increases to reach a maximum and then decreases to zero [30]. This is equivalent to increasing the hopping integrals and thereby the possibility to enhance intraband interactions, which are especially important for the Fe-As-Fe hopping and crucially influence the magnetic properties of the undoped compounds. These are especially sensitive to the As z-axis location and vary up to 10% if the As coordinate is changed by only $\Delta z=0.005$ [31]. Interestingly, however, the ratio of the nearest to next nearest neighbor exchange constants remains the same for as many as 8 structurally different systems with $T_c$'s varying between 18 and 55K. This finding [32] questions that magnetic fluctuations are essential in the pairing mechanism.

The above situation provides consistent explanations for various experimentally observed pecularities of Fe-As based superconductors. In addition, we have presented precise predictions regarding the isotope effect on $T_c$, which enable the identification of the pairing mechanism together with a temperature dependent analysis of the gap behavior, from which the order parameter symmetry can be determined. All results can be tested experimentally and can provide detailed insight into the as yet unidentified origin of superconductivity in Fe-As based compounds.

To conclude, we have presented a detailed analysis of a two gap superconductor with s-wave order parameters. We have shown that a *universal* scaling behavior of the normalized gaps is realized. Several possible origins of isotope effects on $T_c$ have been investigated and it has been shown that doping, respectively $T_c$, dependent isotope effects are possible, if polaronic effects are important. In a BCS type scenario of a two-band superconductor this doping/$T_c$ dependence is negligible but with weakened isotope effect as compared to a single band superconductor. Our results are intended to provide a basis for experimental comparisons. If experimental data strongly deviate from our predictions, it can be concluded that more complex pairing symmetries and interactions are involved.

**Acknowledgement** It is a pleasure to acknowledge stimulating discussions with and a critical reading of the manuscript by K. A. Müller.